\documentclass[twocolumn,aps]{revtex4}

\usepackage{graphicx}

\begin{document}

\title{Gapped spin liquid states in a one-dimensional Hubbard model
with antiferromagnetic exchange interaction}

\author{
Jianhui Dai $^{1}$, Xiaoyong  Feng$^{1}$, Tao Xiang$^{2,3}$, and
Yue Yu$^{2}$ }

\address{$^1$Zhejiang Institute of Modern Physics, Zhejiang University,
Hangzhou 310027,  China }

\address{$^2$Institute of Theoretical Physics, Chinese Academy of Sciences,
P O Box 2735, Beijing 100080, China }

\address{$^3$Interdisciplinary Center of Theoretical Studies,
 Chinese Academy of Sciences,
P O Box 2735, Beijing 100080, China }

\date{\today}

\begin{abstract}

We study the phase diagram of a one-dimensional extended Hubbard
model with antiferromagnetic exchange interaction analytically and
numerically. The bosonization and transfer-matrix renormalization
group methods are used in the corresponding coupling regimes. At
half-filling, the system is a Mott insulator with a finite spin
excitation gap if the on-site Coulomb repulsion is fairly smaller
than the antiferromagnetic exchange $J$. This Mott-insulator is
characterized by the bond-charge-density-wave order or
spontaneously dimerization. In the weak-coupling regime where the
spin-charge separation holds approximately, the critical point
separating the gapless and gapped spin liquid phases is $U_c\sim
J/2$. However, as $J$ increases, the spin-charge couplings become
important and the critical point $U_c$ is significantly suppressed
and eventually tends to zero as $J\to \infty$. Away from
half-filling, the charge gap completely collapses but the spin gap
persists.

\end{abstract}


\maketitle

\section{Introduction}

The notion of spin liquid state was introduced thirty years ago
when P. Fazekas and P.W. Anderson first postulated that due to the
frustrated antiferromagnetic coupling the Mott insulating state on
triangular lattice does not break the spin-rotational
symmetry\cite{Anderson1}. The discoveries of high temperature
superconductivity and other novel correlated many-body phenomena
have stimulated wide interest in spin liquids both theoretically
and experimentally\cite{Anderson2}. For correlated electrons in
two-dimensional Cu-O plane, the most two important energy scales
are the kinetic energy and the on-site Coulomb repulsion. However,
as far as the magnetic properties of a Mott insulator is
concerned, the nearest neighbor antiferromagnetic exchange do play
an important role\cite{Hsu}. Recently, F.C. Zhang showed that
within the Gutzwillar approximation, a quantum phase transition
from a Mott insulator to a gossamer superconductor(which was first
proposed by R.B. Laughlin\cite{Laughlin}) at half-filling can take
place as the on-site Coulomb repulsion is reduced \cite{Zhang}. It
was further anticipated that away from half-filling, this gossamer
superconducting state can evolve smoothly into the
resonant-valence-bond (RVB) spin-liquid phase.

A basic model for investigating the RVB or gossamer states is the
so-called t-U-J model\cite{Hsu,Zhang,Yu,Gan,Chapline}. It is an
extended Hubbard model by explicitly including an
antiferromagnetic exchange interaction. While in two dimensions it
remains challenging to accurately solve this model, it may be
instructive to study the corresponding one-dimensional system.

The one dimensional t-U-J model is described by the following
Hamiltonian
\begin{eqnarray}
H &=&-t\sum_{i\sigma }\left( c_{i\sigma }^{\dagger }c_{i+1\,\sigma
}+h.c.\right) +U\sum_{i}n_{i\uparrow }n_{i\downarrow }
\nonumber \\
&&+J\sum_{i}{\bf S}_{i}\cdot {\bf S}_{i+1}. \label{tuj}
\end{eqnarray}
This Hamiltonian is not merely a toy model motivated by the
corresponding two dimensional system, it is also relevant to
quasi-one-dimensional correlated physics. For example, most of the
Bechgaard salts (the (TMTSF)$_2$X family of quasi-one-dimensional
conductors) show close proximity of spin-density-wave(SDW),
spin-Peierls, ferromagnetic, and superconducting phases by varying
pressures\cite{Bechgaard}. To investigate the coexistence of
triplet superconductivity and ferromagnetism in a class of
quasi-one-dimensional materials, Japaridze et al studied the model
for ferromagnetic exchange with easy-plane anisotropy in the
large-bandwidth limit \cite{Japaridze}. As a byproduct, they
predicted a transition to the dimerized ordering phase in the case
of weak anisotropy by assuming spin-charge separation in the
weak-coupling regime. Comparatively, what was less understood is
the generic feature of the  phase diagram in the U-J plane for the
isotropic antiferromagnetic model(\ref{tuj}).

At the first glance, the situation with generic $U, J>0$ might be
trivial. Because in the atomic limit there is no explicit
frustration between the pair-wise interactions $U$ and $J$, the
system is simply a Mott insulator with gapless spin excitations.
This is in contrast to another well-known one-dimensional model
system, i.e., the conventional extended Hubbard model(CEHM) with
the on-site($U$) and nearest neighbor site($V$) Coulomb
interactions. In the CEHM, the pair-wise interactions $U$ and $V$
are frustrated explicitly, and in the atomic limit, the system is
a charge-density-wave(CDW) insulator for large $V$ and a SDW
insulator for large $U$\cite{Bari}. However, if the on-site
repulsion $U$ is reduced, the virtual hopping processes may affect
the spin physics in Mott insulators.  For example, the recent
studies have shown that even within the CEHM there exists a new
phase in between the CDW and SDW phases. The new phase is
characterized by the bond-charge-density-wave(BCDW) order or
spontaneous dimerization with gapped charge and spin excitations
in a narrow window extending from the weak to intermediate
coupling regimes\cite{Nakamura,Sengupta,Tsuchiizu,note}. Usually,
such an insulating phase is caused by the electron-phonon coupling
or explicit frustrations, but now it is caused by the electronic
correlations only. With this kind of BCDW phase in mind, we are
going to examine the phase diagram of the model (1) in order to
clarify the competing Mott insulator physics in one dimension if
the double occupation is allowed.

In the present work, we study the phase diagram of the half-filled
one-dimensional t-U-J model by using analytical and numerical
methods, depending on the ratio $J/t$. It is found that the ground
state is always a Mott insulator at half filling and a metal away
from half filling. However, the spin excitations have been
dramatically changed by the exchange term. There is a gapped spin
liquid phase characterized by the BCDW order when the exchange
interaction is fairly stronger than the Coulomb repulsion at
arbitrary filling. The spin gap stems from the interplay between
the kinetic energy and the antiferromagnetic exchange interaction.
As spin-charge couplings increase and  the bandwidth decreases,
the gapped spin liquid phase will be suppressed. All these reveal
a new scenario for the creation of spin gaps in a single itinerant
electron chain with translational and spin rotational symmetries
and without explicit frustrations \cite{Oshikawa}.

\section{Weak-coupling theory: bosonization, renormalization group equations,
and semi-classical analysis} Let us first consider the phase
diagram at half filling in the weak-coupling regime $U/t,J/t\ll
1$. In this regime, the bosonization technique may be used
reliably\cite{Emery,Solyom,Voit1,Gogolin}. We linearize the
spectrum and pass it to continuum limit by substituting
$a^{-1/2}c_{j\sigma}\rightarrow (2\pi
a)^{-1/2}\sum_{r=+,-}e^{irk_Fx+ir\varphi_{r,\sigma}(x)}$ where
$\varphi_{r,\sigma}(x)$ are the right/left-moving bosonic fields.
Introducing charge and spin bosonic fields, $\phi_{c,r}=
(\varphi_{r,\uparrow}+\varphi_{r,\downarrow})/2$, $\phi_{s,r}=
(\varphi_{r,\uparrow}-\varphi_{r,\downarrow})/2$, respectively,
the Hamiltonian density of the bosonized model is then given by
${\cal H}={\cal H}_{c}+{\cal H}_{s}+{\cal H}_{cs}$. Here, the
charge and spin sectors are described by
\begin{eqnarray}
{\cal H}_{c}=&&\frac{v_{c}}{2\pi}\sum_{r=+,-} (\partial_x
\phi_{c,r})^2 + \frac{g_{\rho}}{2\pi^2a^2}(\partial_x
\phi_{c,+})(\partial_x
\phi_{c,-})\nonumber\\
&&-\frac{g_{c}}{2\pi^2a^2}\cos 2\phi_{c}\label{hc} \\
{\cal H}_{s}=&&\frac{v_{s}}{2\pi}\sum_{r=+,-} (\partial_x
\phi_{s,r})^2 - \frac{g_{\sigma}}{2\pi^2a^2}(\partial_x
\phi_{s,+})(\partial_x
\phi_{s,-})\nonumber\\
&&+\frac{g_{s}}{2\pi^2a^2}\cos 2\phi_{s}\label{hs}
\end{eqnarray}
with $\phi_{\nu}=\phi_{\nu,+}+\phi_{\nu,-}$ for $\nu=c,s$
respectively. While, the spin-charge coupling part is given by
\begin{eqnarray}
{\cal H}_{cs}=-&&\frac{g_{cs}}{2\pi^2a^2}\cos 2\phi_{c}\cos
2\phi_{s}\nonumber \\-&&\frac{g_{\rho s}}{2\pi^2}(\partial_x
\phi_{c,+})(\partial_x \phi_{c,-})\cos 2\phi_{s}\nonumber\\
+&&\frac{g_{c\sigma}}{2\pi^2}(\partial_x \phi_{s,+})(\partial_x
\phi_{s,-})\cos 2\phi_{c}\nonumber\\
+&&\frac{g_{\rho\sigma}}{2\pi^2}a^2(\partial_x
\phi_{c,+})(\partial_x \phi_{c,-})(\partial_x
\phi_{s,+})(\partial_x \phi_{s,-})\label{hcs}
\end{eqnarray}
Here, in order to compare the t-U-J model with the CEHM, we follow
the weak-coupling $g$-ology approach and adopt the notation of the
Ref.\cite{Tsuchiizu}. In the lowest orders of $U$ and $J$, the
scattering matrix elements are
$g_{1||}=-aJ/2$,$g_{1\perp}=a(U-J/2)$,
$g_{2||}=aJ/2$,$g_{2\perp}=a(U+J/2)$,
$g_{3||}=-aJ/2$,$g_{3\perp}=a(U+3J/2)$, and
$g_{4||}=aJ/2$,$g_{4\perp}=a(U-3J/2)$. The renormalized velocities
and Luttinger couplings of charge and spin sectors are
$v_{c}=2ta+(g_{4||}+g_{4\perp}-g_{1||})/2\pi$,
$v_{s}=2ta+(g_{4||}-g_{4\perp}-g_{1||})/2\pi$  and
$g_{\rho}=g_{2\perp}+g_{2||}-g_{1||}$,
$g_{\sigma}=g_{2\perp}-g_{2||}+g_{1||}$, respectively. The
coupling constants $g_c$ and $g_s$ denote the amplitude of the
backward and the Umklapp scattering of opposite spins, given by
$g_{c}=g_{3\perp}$,$g_{s}=g_{1\perp}$ respectively. $g_{\rho
s}$/$g_{\rho \sigma}$ (and $g_{cs}$/$g_{c\sigma}$) come from the
backward (and Umklapp) scatterings  of the parallel/opposite
spins, respectively, given by $g_{\rho s}=g_{\rho
\sigma}=g_{cs}=g_{c\sigma}=-J/2$ to the lowest order in $J$.

The low-energy properties of the one dimensional t-U-J model in
the weak-coupling regime depend on the scaling behavior of these
coupling constants during the scaling $a\rightarrow ae^{dl}$. Tha
is, the coupling constants run in terms of the following one-loop
renormalization group(RG) equations
\begin{eqnarray}
\lambda dg_{\rho}/dl&=&+2g_{c}^2+g_{cs}^2+g_{s}g_{\rho s}\label{grho}\\
\lambda dg_{c}/dl&=&+2g_{\rho}g_{c}-g_{s}g_{cs}-g_{cs}g_{\rho s}\label{gc}\\
\lambda dg_{s}/dl&=&-2g_{s}^2-g_{c}g_{cs}-g_{cs}^2\label{gs}\\
\lambda dg_{cs}/dl&=&-2g_{cs}+2g_{\rho}g_{cs}-4g_{s}g_{cs}\nonumber\\
&&-2g_{c}g_{s}
-2g_{c}g_{\rho s}-4g_{cs}g_{\rho s}\label{gcs}\\
\lambda dg_{\rho s}/dl&=&-2g_{\rho
s}+2g_{\rho}g_{s}-4g_{c}g_{cs}\label{grhos}\nonumber\\
&&-4g_{cs}^2-4g_{s}g_{\rho s}
\end{eqnarray}
where $\lambda=4\pi ta$. Notice that the SU(2) symmetry in the
spin sector ensures $g_{\sigma}=g_{s}$, $g_{cs}=g_{c\sigma}$, and
$g_{\rho s}=g_{\rho\sigma}$. From these RG equations, one finds
that $g_{\rho}$,$g_{c}$ and $g_{s}$ are marginal with the scaling
dimension 2, while $g_{cs}$ and $g_{\rho s}$ are irrelevant with
the scaling dimension 4. So, we may first neglect ${\cal H}_{cs}$,
assuming spin-charge separation in the weak-coupling regime. In
this case, one obtains the RG flows of $g_{\rho}$, $g_{c}$ from
$\lambda dg_{\rho}/dl=2(g_{c})^2$ , $\lambda
dg_{c}/dl=2g_{\rho}g_{c}$ and $g_{s}$ from $\lambda
dg_{s}/dl=-2(g_{s})^2$, respectively. Since both $U$ and $J$ are
positive and $g_{\rho}(0)=U+3J/2>0$, $g_{c}$ is always relevant
and flows to the strong-coupling fixed point $g_{c}(l)\rightarrow
\infty$. This indicates the existence of the charge excitation gap
at half-filling. The charge gap is  given approximately by $\Delta
_{c}\approx t|g_{c}/\lambda |^{\lambda /2g_{\rho }}$ when $
g_{c}\ll \lambda $. For the spin excitations, $g_{s}$ will flow to
$0$ if $g_{s}>0$, or to $ -\infty $ if $g_{s}<0$ \cite{Gogolin}.
The latter case occurs when $J>2U$ and in this case a spin gap is
expected to open. The spin gap is given approximately by $\Delta
_{s}\approx t\exp \left( \lambda /2g_{s}\right) $ when
$0<-g_{s}\ll \lambda $.

Now, we examine how the spin-charge coupling part ${\cal H}_{cs}$
affects the phase boundary. As $U$ and $J$ increase, the $g_{cs}$
coupling first becomes less irrelevant in terms of the one-loop RG
equations (\ref{grho})-(\ref{grhos}). However, because $g_{c}$
grows with increasing $l$ and dominates over the other couplings,
the charge excitations are always gapful. Below the charge-gap
energy scale the $\phi_{c}$ field is locked at $\langle \cos
2\phi_{c}\rangle\simeq (\Delta_{c}/t)^{2(1-g_{\rho}/\lambda)}$.
Thus, we may re-scale the cosine term in spin sector, by
introducing $g^*_{s}=g_{s}-g_{cs}$. The one-loop RG equation for
$g^*_{s}$ is obtained simply by $\lambda
dg^*_{s}(l)/dl=-2(g^*_{s}(l))^2$. This implies that the spin-gap
transition line is shifted to $g^*_{s}=0$. To obtain the phase
transition line, we follow the same strategy introduced in the
Ref.\cite{Tsuchiizu} and solve
eqs.(\ref{grho})-(\ref{grhos})numerically by looking at which of
the couplings $g_{c}$, $g_{s}$ and $g_{cs}$ becomes relevant, In
our case, $g_{c}$ grows with increasing $l$ faster than the
others. We stop the integration once $g_{c}$ reaches 1 and
calculate $g^*_{s}=g_{s}-g_{cs}$.  The positive(negative)
$g^*_{s}$ leads to the spin gapless(gapful) state. Here, the
vertex corrections to the scattering matrices are not included.
Although they are indeed crucial for the presence of the narrow
BCDW phase in the CEHM\cite{Tsuchiizu}, their effects are limited
in the weak-coupling regime only and may be negligible in the
t-U-J model.  The phase boundary obtained by this way is shown by
the solid line in Fig.1.
\begin{figure}[ht]
\includegraphics[width = 8cm]{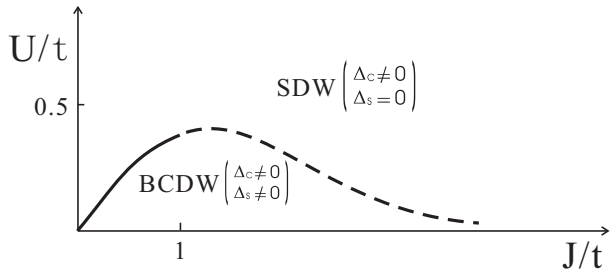}
\caption{The ground-state phase diagram of the t-U-J model at half
filling. The solid line is the phase boundary determined by
numerically solving the RG equations which are valid in the
weak-coupling regime. From the intermediate- to strong-coupling
regimes, the phase boundary is continued by the dashed line which
is qualitatively supported by the numerical data of finite-size
small cluster. The low energy correlations are dominated by the
SDW and BCDW orders in the gapless and gapped spin liquid phases,
respectively. } \label{phase}
\end{figure}

To characterize the gapped spin liquid phase, we examine via a
quasi-classical analysis for four kinds of the order parameters,
the SDW, CDW, BCDW, and bond-SDW(BSDW) parameters, defined by
${\cal O}_{SDW}\equiv (-1)^j(n_{j,\uparrow}-n_{j,\downarrow})$,
${\cal O}_{CDW}\equiv (-1)^jn_{j}$, ${\cal O}_{BCDW}\equiv
(-1)^j\sum_{\sigma}(c^{\dagger}_{j,\sigma}c_{j+1,\sigma}+h.c.)$,
and ${\cal O}_{BSDW}\equiv
(-1)^j(c^{\dagger}_{j,\uparrow}c_{j+1,\uparrow}-
c^{\dagger}_{j,\downarrow}c_{j+1,\downarrow}+h.c.)$. Upon
bosonization, they are written in terms of bosonic fields as
${\cal O}_{CDW}(x)\propto \sin\phi _c \cos\phi _s$, ${\cal
O}_{SDW}(x)\propto \cos\phi _c \sin\phi _s$, ${\cal
O}_{BCDW}(x)\propto \cos\phi _c \cos\phi _s$, and ${\cal
O}_{BSDW}(x)\propto \sin\phi _c \sin\phi _s$ respectively.
Neglecting the spatial variations of the fields we focus on the
following effective potential
\begin{eqnarray}
V_{eff}(\phi_c,\phi_s)&=&-{\tilde g}_c\cos 2\phi _c+{\tilde
g}_s\cos 2\phi _s\nonumber\\
&&-{\tilde g}_{cs}\cos 2\phi _s\cos 2\phi _c .
\end{eqnarray}
The couplings ${\tilde g}$ are all effective ones obtained by
integrating out high-energy degrees of freedom.  As the charge gap
exists everywhere at half-filling and is always large than the
spin gap(${\tilde g}_c$ increases faster than ${\tilde g}_s$ ),
both the BSDW and CDW orders are unfavorable in the ground state.
So in the insulating phase, the SDW and BCDW orders compete each
other, with the effective potential $V_{SDW}=-{\tilde
g}_{c}-{\tilde g}_{s}+{\tilde g}_{cs}$ and $V_{BCDW}=-{\tilde
g}_{c}+{\tilde g}_{s}-{\tilde g}_{cs}$ respectively. It is the
SDW(or BCDW) which dominates in the ground state for ${\tilde
g}_{cs}<{\tilde g}_{s}$( or ${\tilde g}_{cs}>{\tilde g}_{s}$).
Therefore, the SDW-BCDW transition line determined by
quasi-classical analysis is the same as that of the spin-gap
transition line $g^*_{s}=0$ given above. In Fig.(1), the deviation
of the phase boundary from the straight line $U=2J$ with
increasing $J$ is due to the $g_{cs}$ coupling which in turn
enhances SDW order.

\section{Intermediate and strong-coupling regimes}

We believe that the existence of the spin gap is a generic feature
of the t-U-J model, not limited in the weak-coupling regime only.
To further clarify the physics behind the spin-gap, it is
interesting to consider the unconstrained t-J model, namely the
t-U-J model with $U=0$, in the strong coupling limit $J\gg t$. In
this case, one can first bosonize the Heisenberg exchange term and
taking the hopping term as a perturbation. In the pure Heisenberg
model, the spin excitation is critical but the charge excitation
is completely suppressed. Nevertheless, the charge field $\phi
_{c}$ should be introduced in the bosonization of the Heisenberg
term since charge fluctuations are now allowed. For the pure
Heisenberg model, $\phi _{c}$ is pinned to $0$. However, by
introducing a small but finite $t$-term, $\phi _{c}$ will become
finite. This will reduce the charge gap and at the same time
induce effectively a backward scattering term in the spin sector.
As a result of this backward scattering, a spin gap will open.

The discussions in the preceding section and the previous
paragraph showed the existence of the spin gap in both the weak
and strong coupling limits in the unconstrained t-J model. To
examine whether this is true in the intermediate coupling regime
$J\sim t$, we have calculated numerically the zero-field spin
susceptibility using the transfer-matrix renormalization group
(TMRG) at half-filling. The TMRG handles directly an infinite
lattice system. It does not have any finite lattice size effect
and allows a small excitation gap to be accurately determined
without invoking the finite size scaling\cite{Wang}. This is an
advantage and the reason for us to use this method here. The error
comes mainly from the truncation of basis states in the TMRG
iteration. Fig. 2 shows the TMRG result of the uniform spin
susceptibility for $J=t$ and $U=0$. In the TMRG calculation, 100
basis states were retained and the maximum truncation error is
less than $10^{-5}$. As can be clearly seen from the inset of the
figure, the spin susceptibility drops exponentially at low
temperatures. This exponential decay of the susceptibility is a
direct consequence of the spin gap.
\begin{figure}[ht]
\includegraphics[width = 8cm]{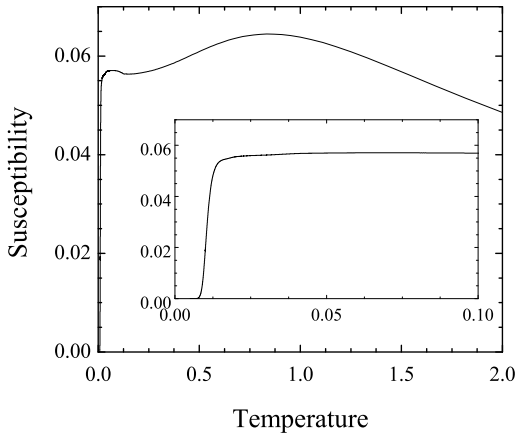}
\caption{Zero field magnetic susceptibility of the one dimensional
t-U-J model with $ J=t=1 $ and $U=0$. The inset is an enlarged low
temperature plot.} \label{suscep}
\end{figure}

So far, we have established the existence of the gapped spin
liquid phase in the weak/strong-coupling regimes and at a special
point in the intermediate regime. Now we try to see the generic
feature of the phase boundary in the U-J plane. As shown in Fig.1,
the phase boundary is very closed to the line $J=2U$ in the
weak-coupling regime, but goes below it as $J$ increases, due to
the spin-charge couplings. We anticipate that the critical line
develops a maximum in the intermediate regime, and then approaches
zero gradually as $J$ goes to infinity, see the dashed line in
Fig.1.

To support this picture and to estimate the crossing point, we
have also studied the spin-gap transition by investigating
level-crossings of the excitation spectra\cite{Julien,Okamoto}.
This technique was used to determine the phase boundaries of the
CEHM by Nakamura\cite{Nakamura} with high accuracy from the
numerical data of finite-size clusters. The obtained spin-gap
transition line is qualitatively in agreement with Fig.1, the
crossing point is about (1.3,0.35)\cite{Feng}.

\section{Away from half-filling: weak-coupling analysis}

Finally, we briefly address the subsequences after introducing
holes into the system. In the weak-coupling regime and away from
half filling, the terms resulting from the Umklapp scattering in $
{\cal H}_{c}$ and ${\cal H}_{cs}$ disappear. The other terms are
still present but their coupling constants become doping
dependent. To the leading order approximation, $g_{\rho
}=a(U+\frac{3J}{2}\cos (\frac{\pi}{2}\delta) )$, $
g_{s}=a(U-\frac{J}{2}\cos (\frac{\pi}{2}\delta) )$ and $g_{\rho
s}=-a\frac{J}{2}\cos (\frac{\pi}{2}\delta) $, where $\delta $ is
the doping concentration. Due to the absence of the Umklapp
scattering, the charge gap collapses. However, the backward
scattering of spin excitations and the subsequent spin gap phase
persist. In the limit $U\rightarrow 0$ and $J\rightarrow 0$, the
spin gap phase exists when the condition $g_{s}<0$, i.e. $J>2U\cos
^{-1}(\frac{\pi}{2}\delta)$, is satisfied. Thus the critical
exchange constant increases with increasing doping. In contrast to
the half-filling case, the gapped spin state is now metallic. In
this Luther-Emery-type state, the CDW and singlet pairing (SP)
correlations develop asymptotically as $ C_{CDW}(r)\sim a_1
r^{-2}+a_2\cos(2k_Fr)r^{-K_{c}}, C_{SP}(r)\sim a_3r^{-1/K_{c}}$,
with $a_i$ being constants of order of 1. Upon doping $\delta>0$,
$K_c=\sqrt{\frac{1-g_{\rho}/2v_c}{1+g_{\rho}/2v_c}}<1$. Of course,
the CDW correlation is the most dominant one in the Luther-Emery
phase.

\section{Summary}

In summary, we have studied the one-dimensional t-U-J model
analytically and numerically in different coupling regimes. At
half-filling, the system shows two distinct insulating phases: a
gapless spin liquid phase dominated by SDW correlation and a
gapped spin liquid phase dominated by BCDW correlation. The
suggested phase diagram is plotted in Fig.1. In the weak coupling
regime, the phase boundary line is determined by solving the RG
equations. This line is anticipated to develop a maximum in the
intermediate regime and tends to zero at the strong coupling
limit. If $U$ is larger than a critical value (which is estimated
about $\sim 0.35t$, see Ref.\cite{Feng}), the gapped spin liquid
no longer exists, irrespective of the magnitude of $J$. After
doping, the gapped spin liquid phase becomes the Luther-Emery
phase where the charge gap is completely suppressed while the
spin-gap persists.

It is remarkable that in the t-U-J model the spin-gap behavior
resembles to that of the frustrated t-J$_1$-J$_2$
model\cite{Imada,Ogata} but with a quite different mechanism and
the charge gap behavior resembles to that of the Hubbard model but
with an enhanced magnitude. In the t-U-J model, the spin gap
results from the interplay between the antiferromagnetic exchange
and the kinetic energy ( the constraint of no double occupancy is
released). The BCDW insulator driven by this mechanism should have
a large but finite ratio of the charge gap to the spin gap,
$\Delta_{c}/\Delta_{s}$. For the band, Kondo and SDW insulators
this ratio is unit, 1-1.5 and $\infty$, respectively. For the
conventional spin-Peierls systems, this ratio should be very
large. So, the present BCDW phase may be relevant to a class of
quasi-one-dimensional Mott-insulators where the charge gaps are
only several times larger than the spin ones. Particularly, we
expect that the mechanism may be relevant to the Bechgaard salts
such as (TMTSF)$_{2}$PF$_{6}$ where a transition from the
spin-Peierls phase into the SDW phase by changing pressures was
observed\cite{Jerome}.

\section{Acknowledgement}
We acknowledge useful discussions with S.J. Qin, Y.J. Wang, Z.X.
Xu and F.C. Zhang. JD wish to thank L. Yu for the hospitality at
the Interdiscipline Center of Theoretical Studies where this work
was initiated. This work was supported in part by the National
Natural Science Foundation of China and the NSF of Zhejiang
Province.

\end{document}